\documentclass[10pt]{article} 

\usepackage[preprint]{tmlr}

\usepackage{amsmath,amsfonts,bm}

\def\eqref#1{equation~\ref{#1}}

\def\1{\bm{1}}

\DeclareMathAlphabet{\mathsfit}{\encodingdefault}{\sfdefault}{m}{sl}
\SetMathAlphabet{\mathsfit}{bold}{\encodingdefault}{\sfdefault}{bx}{n}

\usepackage{hyperref}
\usepackage{url}
\usepackage{xspace}
\usepackage{booktabs}       %
\usepackage{amsfonts}       %
\usepackage{nicefrac}       %
\usepackage{microtype}      %
\usepackage{xcolor}         %
\usepackage{xspace}
\usepackage{microtype}
\usepackage{amsmath}
\usepackage{graphicx}
\usepackage{subfigure}
\usepackage{booktabs}
\usepackage{enumerate}
\usepackage{enumitem}
\usepackage{multirow}
\usepackage{booktabs}

\hypersetup{
    colorlinks=true,
    linkcolor=blue,
    filecolor=blue,      
    urlcolor=blue,
    citecolor=blue
}

\usepackage[textsize=tiny]{todonotes}
\usepackage{xspace} 
\newcommand{\namel}{\ensuremath{{\rm FastLane}}\xspace}

\title{\namel: Efficient Routed Systems for Late-Interaction \\Retrieval}

\author{\name Ramnath Kumar \thanks{Part of work was done during internship at Google Research, India.} \email ramnathkumar181@gmail.com \\
      \addr UCLA. 
      \AND
      \name Prateek Jain \email prajain@google.com \\
      \addr Google DeepMind.
      \AND
      \name Cho-Jui Hsieh \email chohsieh@cs.ucla.edu \\
      \addr UCLA.}

\def\month{MM}  
\def\year{YYYY} 
\def\openreview{\url{https://openreview.net/forum?id=XXXX}} 

\begin{document}

\maketitle

\begin{abstract}
Late-interaction retrieval models like ColBERT achieve superior accuracy by enabling token-level interactions, but their computational cost hinders scalability and integration with Approximate Nearest Neighbor Search (ANNS). We introduce \namel, a novel retrieval framework that dynamically routes queries to their most informative representations, eliminating redundant token comparisons. FastLane employs a learnable routing mechanism optimized alongside the embedding model, leveraging self-attention and differentiable selection to maximize efficiency. Our approach reduces computational complexity by up to \textsc{30x} while maintaining competitive retrieval performance. By bridging late-interaction models with ANNS, \namel~enables scalable, low-latency retrieval, making it feasible for large-scale applications such as search engines, recommendation systems, and question-answering platforms. This work opens pathways for multi-lingual, multi-modal, and long-context retrieval, pushing the frontier of efficient and adaptive information retrieval.

\end{abstract}

\section{Introduction}
\label{intro}

\begin{figure}[hbt!]
\centering
\includegraphics[width=0.8\linewidth]{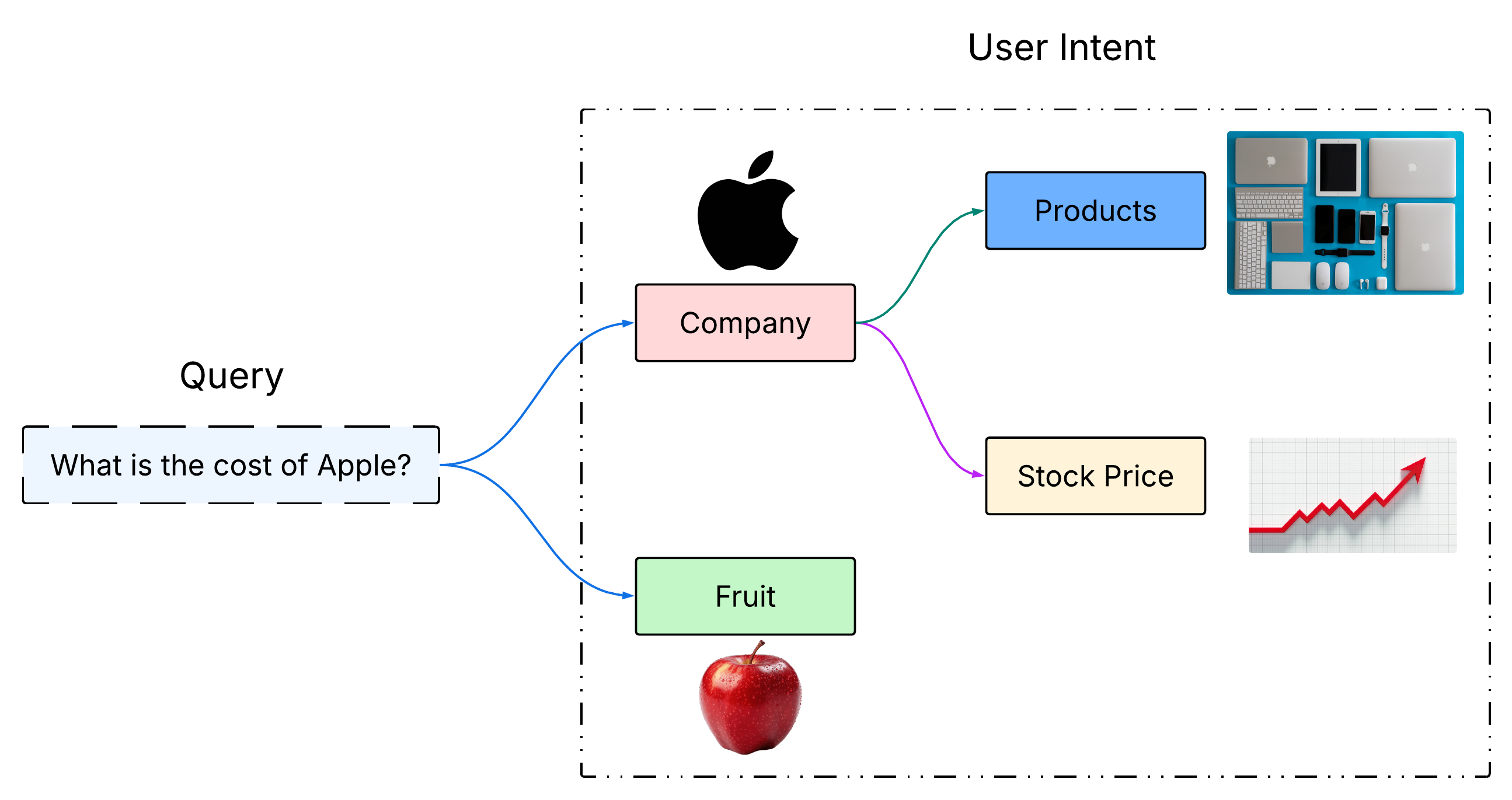}
\caption{The query “What is the cost of Apple?” demonstrates how a single phrase can map to diverse user intents: i.e, from stock prices to fruit depending on context. This underscores the need for fine-grained token-level understanding to disambiguate meaning, especially as conventional NLP preprocessing (e.g., stemming, lemmatization) may obscure critical distinctions by dropping stopwords such as ``a'', ``an'', etc.}
\label{fig:motivation_overview}
\end{figure}

How can we design systems that accurately and efficiently retrieve relevant information while capturing the nuanced meanings of queries? This question lies at the heart of modern information retrieval (IR), a field critical to applications such as search engines, question-answering systems, and recommendation platforms~\citep{johnson2019billion, NayakUnderstanding, dahiya2021deepxml}; retrieval augmented generation~\citep{lewis2020retrieval}; and so much more. Traditional IR methods often rely on single-vector representations of queries and documents. While these approaches are computationally efficient, they frequently struggle to capture the multi-faceted and nuanced semantic relationships present in natural language~\citep{menon2022defense, neelakantan2022text, muennighoff2022sgpt}. For instance, consider the query, ``what is the cost of apple?''; it remains ambiguous whether the question refers to the fruit or one of the various products such as phones and laptops. This example highlights the limitations of single-vector approaches in disambiguating and contextualizing semantic meaning, particularly when queries can have multiple interpretations. This problem arises in many cases of ``branding'', where multiple meanings for the same word, and the true intent of the user would be harder to gauge (see Figure~\ref{fig:motivation_overview}). 

Prior approaches~\citep{khattab2020colbert} have worked towards mitigating these limitations with the use of richer representations. Late-interaction models represent each token with a representation, and use the entire space instead of relying on an arbitrary representation derived from the \textsc{[``CLS'']} token. Late-interaction methods offer a powerful solution to this challenge by representing a ``searched query'' as multiple embedding vectors. State-of-the-Art (SOTA) late-interaction models such as ColBERT~\citep{khattab2020colbert} leverage late interaction (sum-max, or other scoring mechanisms) to derive the relevance score between the query, and the document,  leading to consistently improved performance over  single-view dense retrieval methods~\citep{menon2022defense,neelakantan2022text,muennighoff2022sgpt}.

Late-interaction models require additional computation, and thus are slower, which limits their widespread adoption. This cost in latency arises since late-interaction models require pairwise comparisons between all ``query'' and ``target'' views, instead of the condensed \textsc{[``CLS'']} view alone. 
This hinders the latency of the machine as the late-interaction logic is not parallelizable. This computation limits the scalability, especially for large databases, as the system is no longer compatible with  efficient Approximate Nearest Neighbor Search (ANNS) methods~~\citep{johnson2019billion,sivic2003video,guo2020accelerating}. Thus, most systems in production continue to employ single-view retrieval mechanisms for retrieval, and sometimes take advantage of ColBERT and other late-interaction approaches for re-ranking. Few research works have explored this domain in the scope of dimension size reduction, routing for learning ANNS, but these efforts are still nascent~\citep{dhulipala2024muvera, santhanam2022plaid, ji2024efficient, kumar2023ehi}.

This paper introduces a novel routing-based approach to alleviate this limitation drawing inspiration from this vast literature of methods.  We propose \namel, a class of models upon which late-interaction methods such as ColBERT can now ``pick'' the most informative view for a given query, rather than exhaustively combining information from all views through the \textsc{[``CLS'']} token or other aggregation tricks. This can dramatically reduce computation without sacrificing performance as shown in Section~\ref{sec:results}. Our method learns to dynamically route each query to its most informative representation, enabling seamless integration with ANNS libraries and paving the way for highly scalable, accurate, and efficient multi-view retrieval systems.

\paragraph{Motivation.}

As discussed earlier, the limitation of late-interaction models such as ColBERT involves the non-parallelizable operation. We aim to bridge this gap by enabling the integration of late-interaction models with ANNS, thus improving both retrieval efficiency and result quality. In this work, we explore the extreme setting of ``picking'' a single optimal or informative representation, thus making the approach parallelizable. The implications of such an approach opens up possibilities into serving for longer query context lengths, and better models by default. We propose \namel, a methodology to implement late-interaction models in practice through dynamic view routing. This routing mechanism is trained jointly with the encoder, and leads to an end-to-end differentiable pipeline similar to EHI~\citep{kumar2023ehi}. To the best of our knowledge, \namel~is the first end-to-end learning method for late-interaction dense retrieval search. \namel~is seamlessly compatible with Approximate Nearest Neighbor Search (ANNS).

\paragraph{Overview and Evaluation.}

\namel~leverages the same backbone architecture as ColBERT~\citep{khattab2020colbert} and introduces a self-attention layer~\citep{vaswani2017attention} over the final outputs of the query head. This layer generates logits for ``picking'' each view of the query, which are then transformed into probabilities with the help of the Gumbel-Softmax reparameterization trick~\citep{jang2016categorical}. Furthermore, during training, we take advantage of the straight-through estimator to learn which distinct ``views'' or tokens to route to. Thus, \namel~helps eliminate the summation logic in the late-interaction operation while gradients being backprobable through all the tokens during training. We conduct a comprehensive empirical evaluation of our method against state-of-the-art techniques on standard benchmarks such as MS MARCO~\citep{bajaj2016ms} and TREC-DL 19~\citep{craswell2020overview} (see Appendix~\ref{appendix:datasets}). Our experiments on the MS MARCO benchmark demonstrate that \namel provides a speedup of up to approximately \textsc{30x} while maintaining comparable performance in terms of Mean Reciprocal Rank (MRR@10) and normalized discounted cumulative gain (NDCG@10). Moreover, \namel significantly outperforms previous state-of-the-art single-view dense retrieval approaches by upto \textsc{8.14\%}. We observe a similar trend when working with the TREC DL-19 benchmark.

\paragraph{Contributions.}

This work presents \namel, a novel approach for accelerating late-interaction dense retrieval with dynamic view routing. Our key contributions include:

\begin{itemize}[leftmargin=*]\vspace{-0.1in}
    \itemsep 0pt
    \topsep 0pt
    \parskip 2pt
    \item \textbf{Efficient Late-Interaction Retrieval with ANNS Compatibility.} \namel eliminates the summation operation in late-interaction scoring through a learnable routing function, enabling parallelization and seamless integration with Approximate Nearest Neighbor Search (ANNS). This reduces computational complexity by up to \textbf{30x} while maintaining retrieval effectiveness. (see Section~\ref{sec:ablations})
    
    \item \textbf{Strong Empirical Performance.} We conduct extensive evaluations on the MS MARCO and TREC-DL benchmarks, demonstrating that \namel achieves competitive performance against state-of-the-art (SOTA) methods, including ColBERT, SGPT, ANCE, and DyNNIBAL, while significantly improving efficiency. (see Section~\ref{sec:results}, Appendix~\ref{appendix:extended_results})

    \item \textbf{Model-Agnostic Design.} While implemented with a ColBERT backbone, \namel is adaptable to various encoder architectures, similarity metrics, and hard negative mining strategies, making it broadly applicable across dense retrieval frameworks. (see Section~\ref{sec:background})

\end{itemize}

\section{Related Work}
\label{sec:related_works}

The field of information retrieval has witnessed remarkable advancements, progressing from single-vector representations to more nuanced late-interaction methods that capture the complexities of queries and documents. These developments can be briefly categorized as (i) Single-Vector Retrieval Models, (ii) Late-Interaction Models, (iii) Static Models, and (iv) Dynamic Learnable Models for Routing.

\textbf{Single-Vector Retrieval Models.} Recent advances in contrastive learning~\citep{gutmann2010noise} helped power strong dual encoder-based dense retrievers~\citep{ni2021large,izacard2021towards,NayakUnderstanding}. These dual-tower models consist of query and document encoders, often shared, which are trained with contrastive learning using limited positively relevant query and document pairs~\citep{menon2022defense,xiong2020approximate}. These models condensed semantic information into the [CLS] token, enabling compatibility with Approximate Nearest Neighbor Search (ANNS)~\citep{johnson2019billion}. ANNS facilitated efficient retrieval by clustering the corpus and limiting the search space to smaller, relevant subsets of the original data. However, the simplicity of single-vector embeddings failed to accurately capture the rich, multi-dimensional semantics of complex queries and documents~\citep{neelakantan2022text, muennighoff2022sgpt}. This limitation motivated the development of retrieval systems with token-level granularity, laying the groundwork for such models.

\textbf{Late-Interaction Models.} Late-interaction models, as introduced through ColBERT~\citep{khattab2020colbert}, represent a exploration of a new paradigm of models in retrieval. By introducing the use of token-level embeddings and a \textit{sum-max} aggregation mechanism, these models utilized the ability to handle fine-grained semantic relationships between queries and documents. This approach significantly improved retrieval precision, making late-interaction models well-suited for nuanced information retrieval tasks. Building on this, ColBERTv2~\citep{santhanam2021colbertv2}, SPLADE~\citep{formal2022distillation} enhanced efficiency by refining the late-interaction mechanism and introducing lightweight document representations. However, as discussed earlier, these models came with its own limitations in latency, and memory bottlenecks.

\textbf{Static Model Paradigm.} Due to the high memory, and energy requirements of late-interaction methods, their wide-spread adoption have remained limited. Workarounds in the form of two-stage pipelines helped reduced the load to the late-interaction models by re-ranking to fewer documents. However, ANNS were not designed to adapt to this change in paradigm. Multiple works have explore this dual-stage setup from a pre-trained baseline ColBERT setup. PLAID~\citep{santhanam2022plaid} introduced a multi-stage pipeline, progressively narrowing down candidate documents to minimize computational overhead. 
Along tangential lines, MUVERA~\citep{dhulipala2024muvera} transformed multi-vector similarity search into single-vector similarity search through Fixed Dimensional Encodings (FDEs), improving compatibility with ANNS solvers while maintaining retrieval accuracy. However, the cost of retrieving fewer views came with an increased dimension size, which increases the ANNS complexity linearly, while the number of views only increased the latency in a logarithmic fashion (as discussed briefly in Section~\ref{sec:background}). These approaches demonstrated the feasibility of scaling token-level interactions to larger datasets and complex retrieval scenarios, but remained disjoint steps.

\textbf{Dynamic Model Paradigm.} As retrieval tasks grow in complexity, trainable/learnable scoring models were introduced in Lite ColBERT~\citep{ji2024efficient}. Lite ColBERT~\citep{ji2024efficient} combined the sum-max aggregation logic to a single attention scorer, followed by an MLP for representing the final similarity score. The Lite model helps reduce the number of tokens overall, making late-interaction models more practical during optimization. However, the aggregation would still require us to build a new paradigm of ANNS models for the task. We believe End-to-End Hierarchical Index (EHI)~\citep{kumar2023ehi} is the closest work in this literature which works with routing as an effective strategy to scale dense retrieval systems. By structuring the search space hierarchically, EHI significantly reduces computational costs, making them highly compatible with large-scale ANNS frameworks. 

On a parallel effort, we introduce \namel, a retrieval paradigm of models that bridges the efficiency of dual-tower models with the accuracy of late-interaction methods. By employing a differentiable routing mechanism optimized alongside the embedding model, \namel~dynamically identifies the most informative representations for retrieval. This selective routing reduces computational complexity by 30x (evaluated on queries with up to 30 tokens) while maintaining competitive accuracy. Moreover, \namel~seamlessly integrates with off-the-shelf ANNS frameworks and dynamic frameworks such as EHI~\citep{kumar2023ehi}, providing a versatile solution for both existing and emerging retrieval systems. To the best of our knowledge, EHI remains the closest work to \namel, where EHI works on routing to train a retrieval model in an end-to-end fashion. In this work, we extend the paradigm of EHI to bridge late-interaction models and current architectures.
\section{Background}
\label{sec:background}

\paragraph{Problem Definition and Notation.}
Imagine we have a collection of $N$ queries, $\mathcal{Q} = \{q_1, \dots, q_N\}$, and $M$ documents, $\mathcal{D} = \{d_1, \dots, d_M\}$. Each query-document pair $(q_i, d_k)$ has a label $y_{ik}$, where $y_{ik} = 1$ means the document is relevant to the query, and $y_{ik} = -1$ means it is not. The goal is to build a system (called a \textit{retriever}) that can take any query and find the most relevant documents from the collection.

Dense embedding-based retrieval~\citep{johnson2019billion,NayakUnderstanding,dahiya2021deepxml, khattab2020colbert,menon2022defense,neelakantan2022text,muennighoff2022sgpt} is the state-of-the-art (SOTA) approach for semantic search and typically involves embeddings the documents and queries alike using a deep networks like BERT~\citep{devlin2018bert}. 

\textbf{Single-View Approaches.}
Traditional single-view approaches~\citep{menon2022defense,neelakantan2022text,muennighoff2022sgpt} often involve a forward pass through a transformer to obtain a single embedding representation for both the query and document using the CLS token. The objective is to align the query CLS representation with the document CLS representation. Note that $W_s$ is typically a projection matrix used to map the BERT output into a shared embedding space for computing similarity score. In our experiments, we train a series of BERT-based ColBERT models, and regular dual encoder models on the (“small”) triplets training set, employing 6-layer BERT models from \citep{turc2019well}. Assuming a fixed maximum token length of m for queries and n for documents, the similarity score between a given query and document can be calculated as follows:

\begin{equation}
\begin{aligned}
    \hat{q} &= \texttt{BERT}([CLS; q_{1:m}])_1 \cdot W_s, \\
    \hat{d} &= \texttt{BERT}([CLS; d_{1:n}])_1 \cdot W_s, \\
    s &= \hat{q} \cdot \hat{d}.
\end{aligned}
\end{equation}

However, single-view embeddings can struggle with queries that span multiple nuanced interpretations. For instance, consider the query ``what is the cost of the apple?'' This query might relate to the stock price of the company , or the fruit. Despite their obvious semantic difference, the propagation of which information users wish to ask for re-brands the information learned by the models. Thus, a single-view representation may fail to capture these diverse connections, grouping the query closer to a concept while missing the broader context.  The latency to serve for these models via an off-the-shelf ANNS such as SCANN~\citep{guo2020accelerating}, FAISS~\citep{johnson2019billion}, or other hierarchical methods is $\mathcal{O}({n\log d})$, where $n$ is number of dimensions, and $d$ is the number of documents in our corpus. Such bottlenecks would also apply for MUVERA.

\textbf{Multi-View Approaches.} Late-interaction based~\citep{khattab2020colbert} methods which uses the representative power of multiple tokens addresses the single-view approach by accounting for the various semantics. The intuition as we understand it is that given a query, and the documents, those representations extracted from each token (through attention-based mechanisms to account for longer contexts) of the query are deemed relevant for representation learning, and thus retrieval. Assuming a fixed maximum token length of m for queries and n for documents, the similarity score between a given query and document can be calculated as follows:

\begin{equation}
\begin{aligned}
    \hat{q}_i &= \texttt{BERT}([CLS; q_{1:m}])_i \cdot W_s, & \mathcal{Q} &= \langle \hat{q}_0, \hat{q}_1, \dots, \hat{q}_m \rangle, \\
    \hat{d}_i &= \texttt{BERT}([CLS; d_{1:n}])_i \cdot W_s, & \mathcal{D} &= \langle \hat{d}_0, \hat{d}_1, \dots, \hat{d}_n \rangle, \\
    s &= \sum_{i=0}^{m} \max_{d_j \in \mathcal{D}} \langle \hat{q}_i, \hat{d}_j \rangle.
\end{aligned}
\end{equation}

Although the approach aggregates via the Sum-Max late interaction, the increased capacity of the search space, the same also makes the approach slow as it no longer parallelizable. This limits the adoption of these models and adaptable by off-the-shelf indexers which are primarily catered to single-view models. Furthermore, even if we deploy an ANNS for each view of the query, one would have a serving cost of $\mathcal{O}(v_{\text{query}}\cdot n \cdot \log(d\cdot v_{doc}))$, where $v_{\text{query}}$, and $v_{\text{doc}}$ are the number of tokens in the query, and the document respectively. $n$ is the number of documents, and $d$ is the dimension size. As noted from the asymptotic notations above for the ColBERT model, the computation complexity increases linearly in terms of the context-length of the query, and logarithmically in terms of the context-length of the document.

\subsection{Overview of \namel}
\label{sec:analysis}

Before delving further, we provide the intuition behind our approach to ground our findings in a simpler conceptual framework. People naturally possess a unique authorship style, shaped by their experiences and various influencing factors~\citep{cheng2011author}. These styles, while distinct, often adhere to the foundational principles and structures of the language being used, as observed in prior work~\citep{kumar2020edarkfind}. Languages inherently exhibit redundancies, such as stopwords, which act as connectors but do not typically drive the core intention of a query or text. We hypothesize that these redundancies create natural clusters in the token-space, where tokens sharing similar semantic or structural roles are likely to group together. This clustering serves as the basis for our approach.

We analyzed the token views generated by a traditional ColBERT model pre-trained on the LoTTE benchmark (see code \footnote{https://github.com/stanford-futuredata/ColBERT}). We hypothesized that many tokens share semantically similar meanings, given that even with as many as 30 tokens, a typical query likely represents only 4-6 distinct semantic interpretations or views. Our intuition inspired by prior works could provide useful in the domain of retrieval.

\begin{figure}[hbt!]
\centering
\includegraphics[width=0.9\linewidth]{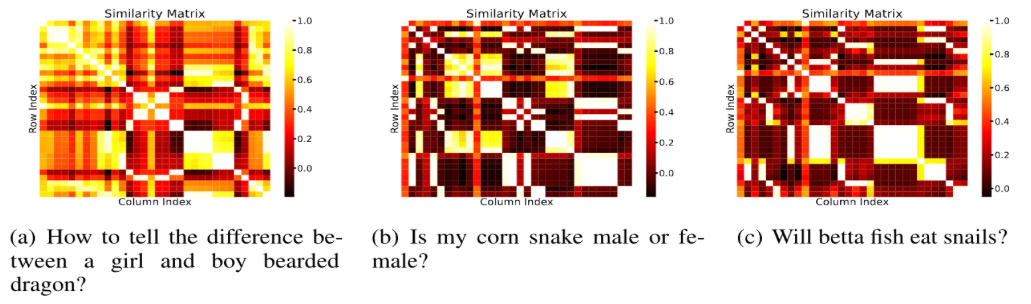}
\caption{Similarity matrices computed from embeddings generated by a pre-trained ColBERT model on the LoTTE benchmark for both query and document examples. The figure illustrates the token-level similarity matrix for a query, highlighting how a small number of distinct token interactions effectively capture the core semantics of the query.
}
\label{fig:motivation}
\end{figure}

As illustrated in Figure~\ref{fig:motivation}, our analysis confirms this intuition, revealing that many token representations exhibit high correlations, resulting in an interesting scenario where few unique clusters remain.  By applying agglomerative clustering with a stringent threshold of $0.95$, we observe very few distinct clusters (approximately 4-6 distinct representations). Consequently, the sum-max operation essentially reduces to a weighted average of a small subset of views. By selectively picking the most common query view through our scoring mechanism, we can bypass the summation process and seamlessly integrate multi-view retrieval with ANNS. 

\subsection{Our Proposed Approach}
\label{sec:motivation}

Given this confirmation in our hypothesis, we aim to design a routing based multi-view search where only a given ``view'' (note that we will now starting to use ``view'', and ``token representation'' interchangably) of query is required to approximate the overall score. Due to redundancy of many of these tokens, we hypothesize that there might be an intrinsic signal to help learn this efficiently. 

\begin{figure*}[hbt!]
\centering
\includegraphics[width=0.9\linewidth]{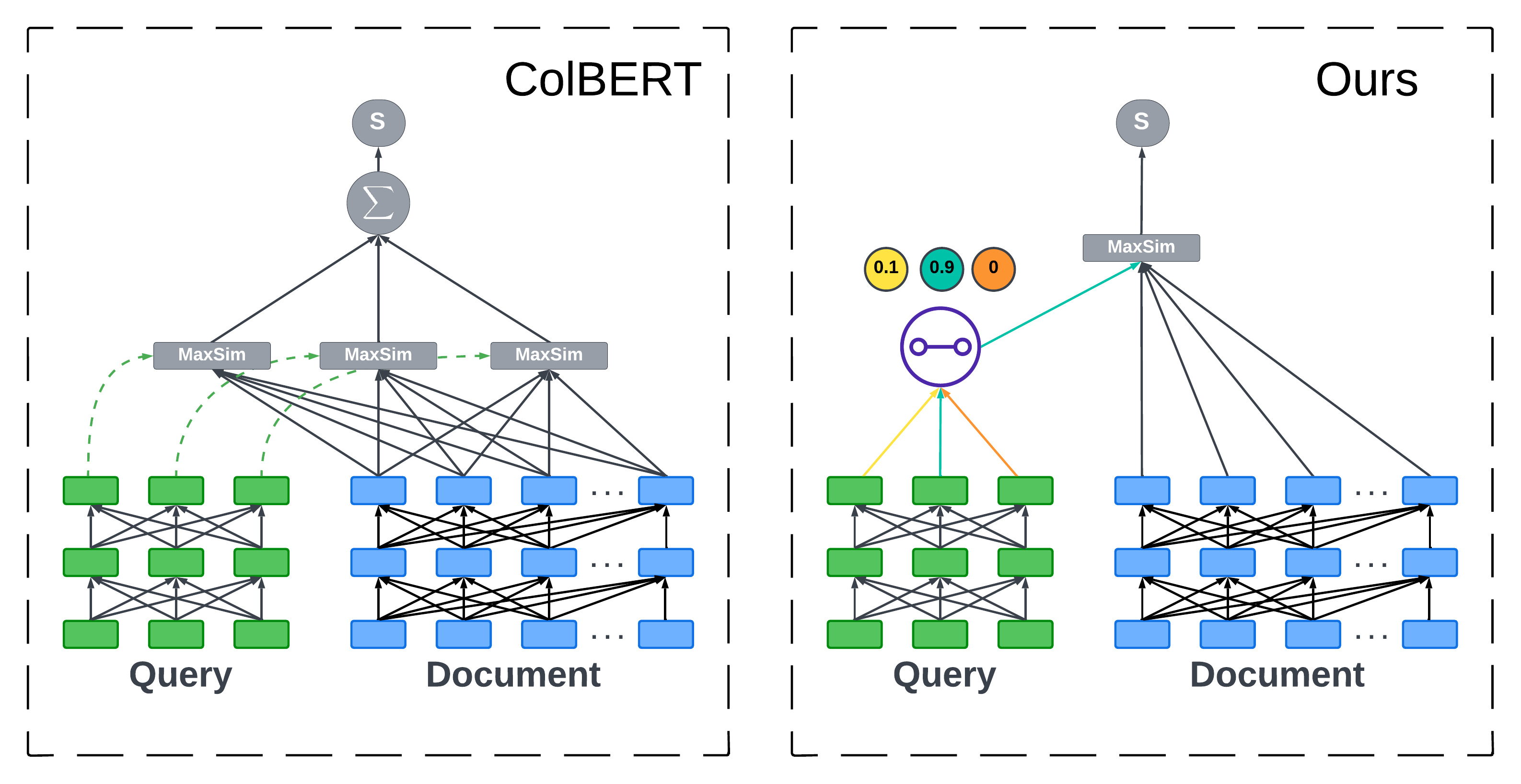}
\caption{Comparison between ColBERT, and \namel paradigm of models.}
\label{fig:compare_baselines_fastlane_idea}
\end{figure*}

Our proposed system introduces a novel mechanism to optimize the retrieval process. Specifically, given the $M$ token embeddings of a query derived from a transformer model, we first pass these embeddings through a Self-Attention Layer followed by a Dense Layer. This pipeline generates a single probability value using a softmax activation function, that allows us to estimate the informativeness of each representation. Intuitively, if a specific token view achieves a higher probability, it signifies that the view captures more critical information about the query-document relationship. To enable efficient learning in this probabilistic framework, we incorporate the Gumbel-Softmax reparameterization trick~\citep{jang2016categorical}, that ensures that the selection process remains differentiable. Additionally, we employ a straight-through estimator to maintain an end-to-end training pipeline, inspired by the approach in EHI~\citep{kumar2023ehi}. This mechanism as proposed in EHI allows us to dynamically select a single representation from the $v_{\text{doc}}$ candidate views of a document during retrieval as well (see Figure~\ref{fig:compare_baselines_fastlane_idea} for an overview). While our current implementation focuses on selecting one view per document, future extensions could explore selecting multiple diverse views, potentially capturing multilingual or multi-modal representations$-$a direction we highlight as an exciting avenue for further research. Our approach introduces minimal overhead to the retrieval process while maintaining computational efficiency. The overall complexity is bounded by $\mathcal{O}(n \cdot \log (d \cdot v_{\text{doc}}))$, where $n$ is the number of dimensions, $d$, is the number of documents in the corpus, and $v_{\text{doc}}$ is the number of views per document which we would need to index. Furthermore, we could also significantly reduce the number of documents to index as depicted in Figure~\ref{fig:motivation}. This would significantly reduce the memory footprint of the model if we could store only few pertinent (capturing most information or diverse views). This reduced memory bottleneck is crucial for serving to consumers at the scale of billions or trillions.

Furthermore, \namel~has a performance cost of $\mathcal{O}(n \cdot \log(d\cdot v_{doc}))$, a significant speedup of $v_{\text{query}} $ over ColBERT (as described in Section~\ref{sec:background}), and maintains performance (as shown in Section~\ref{sec:results}). This  opens up interesting avenues in the domains of multi-lingual retrieval, long-context retrieval, multi-modal retrieval, and the current interests in retrieval augmented generation~\citep{lewis2020retrieval}. 

The FastLane model incorporates a multi-view retrieval strategy that leverages a self-attention mechanism to determine the importance of different views, followed by a probabilistic selection process using the Gumbel softmax reparameterization. The model is trained end-to-end, with a masking approach that regularizes view selection.

\paragraph{Self-Attention Mechanism}

The \namel~model begins by taking multiple embedding representations as returned by the backbone ColBERT model~\citep{khattab2020colbert}. For each view, the model applies a self-attention mechanism~\citep{vaswani2017attention} to compute the relative importance of that view in the context of retrieval. Let the input token embeddings be represented as 
${v_1,v_2,\dots,v_n}$ where each $v_1$ is the representation of the first token, and so on. The self-attention mechanism computes attention scores for each view based on their pairwise interactions, as follows:

\begin{equation}
    A_{ij} = \frac{Q(v_i) \cdot K(v_j)}{\sqrt{d_k}}
\end{equation}

where $Q(v_i)$ and $K(v_j)$ are query and key transformations (restricted to single layer MLP in our case) respectively, and $d_k$ is the dimensionality of the key vectors. The attention scores are then passed through a softmax layer to normalize them across all the tokens $\alpha_{ij} = \text{softmax}(A_{ij})$. This produces a distribution over the different tokens, where each token is learned to approximate the ``relevance to the retrieval task'' as we progress through training.

\paragraph{Gumbel Softmax Reparameterization}

To make the view selection process differentiable, the \namel~model employs the Gumbel-Softmax reparameterization~\citep{jang2016categorical} trick. The Gumbel-Softmax is a continuous approximation of discrete categorical sampling that allows backpropagation through the sampling process. To briefly summarize the process: Initially, a Gumbel noise $g_i$ is added to the logits of the view probabilities $\alpha_i$, which perturbs the logits to encourage exploration $g_i = -\log(-\log(u_i)),$ where $u_i \sim \text{Uniform}(0, 1)$ are random samples from the uniform distribution. The perturbed logits are then passed through the softmax function:

\begin{equation}
    \hat{\alpha}_i = \frac{\exp((\log(\alpha_i) + g_i)/\tau)}{\sum_j \exp((\log(\alpha_j) + g_j)/\tau)},
\end{equation}

where $\tau$ is the temperature parameter controlling the smoothness of the distribution. As $\tau \to 0$, the distribution approaches a one-hot categorical distribution, and mimics picking. Furthermore, we note that other activation functions are also possible in this setting, including but not limited to Gumbel-Tanh, or using diverse sampling schemes such as Determinantal Point Processes (DPP; \citep{kulesza2012determinantal}) but such efforts remain beyond the scope of the current work, and are promising research avenues. 

\paragraph{Straight-Through Estimator}

To further facilitate seamless training and inference with the same architecture (similar to off-policy training, and on-policy evaluation), \namel~uses the straight-through estimator (STE)~\citep{bengio2013estimating}. The STE allows gradients to propagate through the Gumbel softmax by treating the hard one-hot vectors as ``continuous'' during the forward pass and ``discrete'' during the backward pass. This enables the model to maintain end-to-end differentiability while performing discrete view selection. During training, the view with the highest score is selected though the straight-through estimator which picks the arg-max value over the softmax output. In the backward pass, gradients are passed through the continuous softmax outputs instead of the discrete arg-max, ensuring that the gradient flow remains intact.

\paragraph{Masking and Regularization}

The \namel~model incorporates a masking mechanism to help select the view. The mask based on the view probabilities, ensuring that views with less information are masked out during training. The mask is a binary vector $m_i \in {0, 1}$, where a value of 0 indicates that the corresponding view is masked out. To avoid trivial solutions, the mask is applied with a small constant $\epsilon$ that ensures smoother convergence $\alpha_{i} = \alpha_{i} \cdot (1 - m_{i}) + \epsilon.$, and prevents the model from completely ignoring certain views. The added $\epsilon=0.05$ in our models acts as a regularizer, encouraging the model to explore during training.

\paragraph{End-to-End learning}

The model is trained end-to-end.  The self-attention mechanism learns to weight the importance of different views. The Gumbel-Softmax allows for differentiable selection of a single view. The STE enables gradient flow through the discrete selection process. The masking mechanism, combined with the small constant $\epsilon$, regularizes the learning, preventing the model from overly relying on a single view and promoting exploration of different views.  The loss function compares the score of the selected view against the ground truth relevance, driving the model to learn informative view selection for efficient retrieval.

\section{Experiments}
\label{sec:experiments}

In this section, we depict how our proposed approach (\namel) compares to SOTA single-view dense retrieval approaches~\citep{menon2022defense,neelakantan2022text,muennighoff2022sgpt}, and other late-interaction methods such as ColBERT~\citep{khattab2020colbert}. We train a series of BERT-based ColBERT models, and regular dual encoder models on the (“small”) triplets training set, employing 6-layer BERT models from \citep{turc2019well}. We train all models from scratch for 1.5 million steps until convergence. We also use the KL-distillation and obtain slightly better performance from our ColBERT models~\citep{hofstatter2020improving}. We report comparisons against these baselines on the MS MARCO and the TREC DL-19 benchmark as depicted below. Other numbers were taken as is from prior reports of results as collated in \citep{kumar2023ehi, menon2022defense}. Our experimental setup is similar to the setup proposed in~\citep{menon2022defense}.

\subsection{Results}
\label{sec:results}

\begin{figure*}[hbt!]
    \centering
    \includegraphics[width=\linewidth]{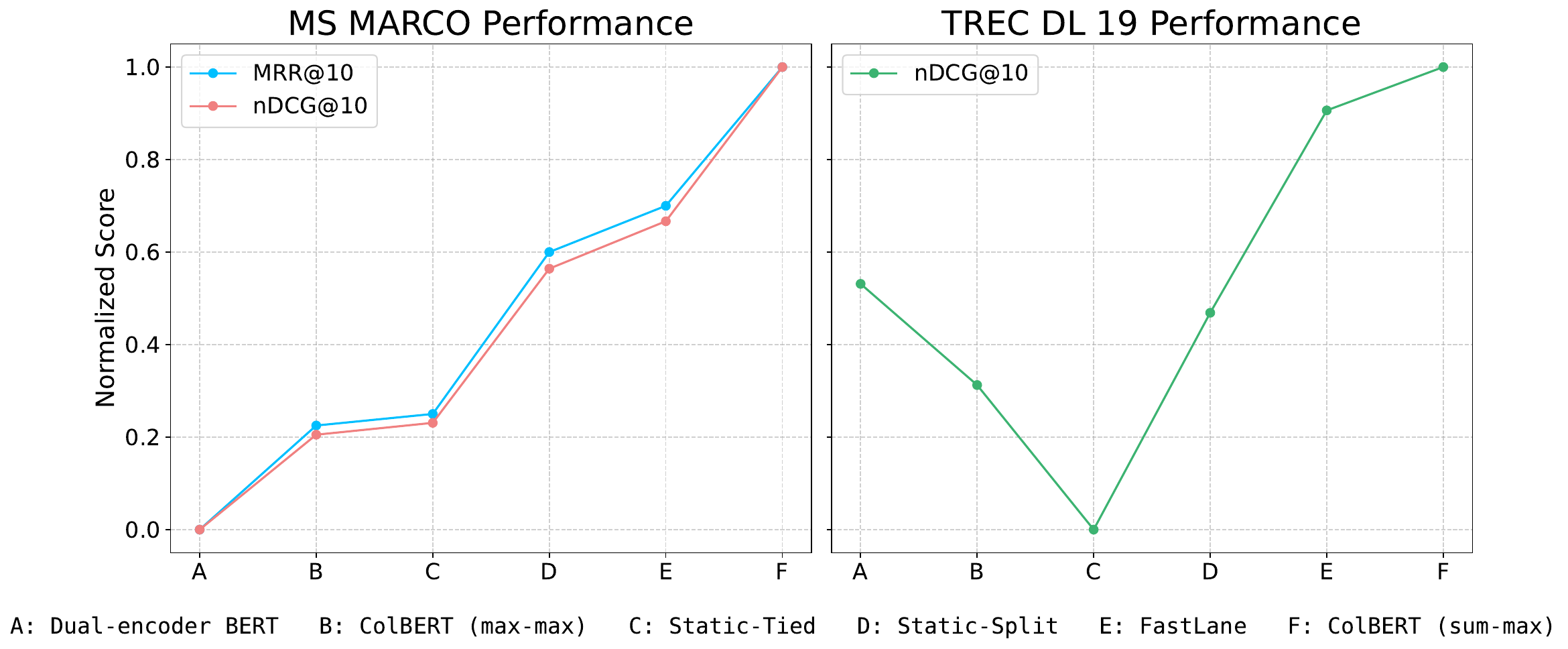} 
    \caption{Relative performance of different models on MS MARCO and TREC DL-19 datasets, showing improvement over the Dual-encoder BERT baseline and normalized between 0 and 1.}
    \label{fig:performance}
\end{figure*}

Figure~\ref{fig:performance} depicts that \namel~is competitive to the ColBERT benchmark while being significantly faster. Appendix~\ref{appendix:extended_results} depicts a more detailed comparison with additional baselines such as SGPT, cpt-text, ANCE, DyNNIBAL, DSI amongst others. Furthermore, parallel research endeavours around negative mining, distillation and others remain an open and interesting avenue for such models. As showcased in Table~\ref{tab:main_comparison}, \namel~achieves competitive performance in comparison to SOTA and is servable at the latency as single-view dual-encoder models as showcased in the table. This is an improvement of approximate \textbf{8.14\%} in MRR@10, \textbf{6.4\%} in nDCG@10 on MS MARCO benchmarks, and \textbf{1.6\%} in nDCG@10 on the TREC DL-19 benchmark as further depicted in Table~\ref{tab:ablations}. This paradigm of learning a routing function for query representations while maintaining the advantages and increased capacity of these multi-view approaches, helps place \namel, as a step towards wider adoption of these models, and interesting research directions towards longer context lengths of queries, multi-lingual retrieval models, multi-modal retrieval models, and many more.

\textbf{Routing Decision.} To understand the effect various routers, we design experiments comparing \namel~against another retrieval algorithms with varying routing intuitions. In this work, the goal has been to choose a single view for the query. Naively, this could be done when one resorts to only using the ``CLS'' token output for query (average of all views without focussing on the relative importance of each of them). This would essentially keep the cost similar to \namel. However, as showcased in Table~\ref{tab:ablations}, \texttt{Static-Tied} (with tied towers where query, and document towers share weights), this is shown to be worse off in comparison to \namel. Furthermore, we create another ablation where we do not tie the towers and do note a significant improvement as depicted in Table~\ref{tab:ablations}, titled \texttt{Static-Split} (with different towers where query, and document towers do not share weights). This could be viewed analogous as using only the [``CLS''] token from the query model, while using all the token representations in the corpus. As showcased in Table~\ref{tab:ablations}, we note that \namel~outperforms these traditional dual-encoder (single-view) models as well as alternative static routing schemes through learning a dynamic routing mechanism. 

\begin{table*}[htbp]
\centering
\caption{Performance metrics of $\text{MRR}@10$ and $\text{nDCG}@10$ evaluated on the MS MARCO dev set and TREC DL-19. The best result for each metric is shown in \textbf{bold}, and the second-best is marked with $\dag$.}
\label{tab:ablations}
\renewcommand{\arraystretch}{1.2}
\setlength{\tabcolsep}{10pt}
\resizebox{0.95\linewidth}{!}{%
\begin{tabular}{lccc}
\toprule
\textbf{Model} & $\textbf{MRR}@10$ (MS MARCO) & $\textbf{nDCG}@10$ (MS MARCO) & $\textbf{nDCG}@10$ (TREC DL-19) \\
\midrule
\texttt{Dual-encoder BERT}       & 0.344 & 0.404 & 0.742 \\
\texttt{ColBERT (max-max)}       & 0.353 & 0.412 & 0.735 \\
\texttt{Static-Tied}             & 0.354 & 0.413 & 0.725 \\
\texttt{Static-Split}            & 0.368 & 0.426 & 0.740 \\
\texttt{FastLane}                & 0.372$^\dag$ & 0.430$^\dag$ & 0.754$^\dag$ \\
\texttt{ColBERT (sum-max)}       & \textbf{0.384} & \textbf{0.443} & \textbf{0.757} \\
\bottomrule
\end{tabular}}
\vspace{1ex}
\end{table*}

\textbf{Latency gains.} Conventional late-interaction models (including ColBERT~\citep{khattab2020colbert}) traditionally use the sum-max aggregation, and are not readily integrable with Approximate Nearest Neighbor Search (ANNS). Thus, these models are not readily deployed in practice. Instead, practitioners often use single-view dual-encoder models in practice, and sometimes rely on ColBERT models for re-ranking. In this work, we proposed a methodology to overcome this two-stage process and combining these steps into a single pipeline. \namel~achieves this by ``picking'' a view of query to use, and eliminating the summation part in our computation. \namel~shows a reduced computation of almost \textsc{30x}. Since the \namel~methodology adds minimal additional parameters to the ColBERT model, computation of embeddings for queries are roughly identical to the traditional ColBERT model. In our experimental setup, the context-length of the query is limited to 30, and the context-length of the document is limited to 200. The embedding size used in each of the representations of these ``views'' is fixed to $768$. With a corpus size of $100,000$ unique documents, we note that the time taken to retrieve documents in traditional late-interaction approaches such as ColBERT. Our analysis on T4 GPUs showcases a significant improvement in retrieval latency: ColBERT takes $112.04$ seconds, whereas \namel reduces this to just $14.48$ seconds, demonstrating nearly an \textsc{8x} speedup. Although, the proposed approach falls short of the traditional ColBERT models by 1\% in terms of performance, we believe that the speedup achieved, and the possible integration of multi-view approaches for retrieval in practice opens up new avenues for research and is of interest to the general community.

\vspace{-1em}


\section{Limitations and Future Work}
\label{sec:ablations}

In this section, we justify the various design choices, and lay the groundwork for few interesting future research endeavors around the paradigm of \namel~models.

\textbf{Memory Bottlenecks} One limitation, that remains unsolved with the current approach is significantly larger the memory footprint of traditional late-interaction approaches such as ColBERT~\citep{khattab2020colbert}. Note that a cost of indexing in traditional single-view models is $d$ documents of the corpus. However, in late-interaction models such as ColBERT, one would need to index $\mathcal{O}(d\cdot v_{\text{doc}})$, where $d$ is the number of documents, and $v_{\text{doc}}$ is the number of document tokens or views. This number is usually large, around 200 tokens in our experiments, but could be much larger as well. Some works including \citep{santhanam2021colbertv2} works upon quantization and reduction of dimensions of the embeddings. We note this as a potential direction to help integrate \namel~with. Another option is to cluster these view embeddings per document and use a Determinantal Point Processes (DPP) or clustering followed by picking diverse views to help reduce memory footprint of our multi-view models as shown in Figure~\ref{fig:motivation}.

\textbf{Future Works.} Through our study as depicted in Table~\ref{tab:main_comparison}, we notice an interesting avenue of future research endeavor where \namel~could facilitate the use arbitrarily longer context-length in the query side. This is further corroborated from our results as depicted in Table~\ref{tab:ablations}. This improved performance of \namel~over traditional single-view dual encoder approaches shows promise not only for enabling more powerful algorithms in our general systems, but also enables the use of arbitrarily longer context length in the query side (adhering to memory and fairness constraints). The multi-view embedding space, even if sparsely accessed, provides a richer representation capacity. This allows for finer-grained semantic distinctions, potentially capturing nuances that the single-view encoder struggles to represent (improving expressiveness, while reducing computational cost). The view selection mechanism acts as a dynamic ``switch'', adapting to different queries and selecting the most suitable subspace within the larger multi-view embedding space. This provides flexibility that a fixed single-view encoder lacks. Such an approach opens up avenues for research of integrating with CLIP~\citep{radford2021learning} to account for multiple languages and facilitate significantly longer query context lengths for retrieval. This would have direct implications on Multi-lingual retrieval, Multi-modal retrieval, and  Retrieval Augmented Generation~\citep{lewis2020retrieval}. 

\section{Conclusion}

In this work, we introduce \namel, a learnable routing paradigm that enhances late-interaction models like ColBERT by balancing accuracy and efficiency. By dynamically scoring and selecting the most informative token-level representations, \namel~significantly reduces the computational complexity of multi-view retrieval systems. This enables \namel~to bridge the gap of late-interaction models and off-the-shelf ANNS, making it a scalable and practical solution for large-scale retrieval tasks. \namel~has an intrinsic ability to handle longer context lengths efficiently for retrieval. By improving the scalability, adaptability, and computational efficiency of these late-interaction models, \namel~represents a shift towards a new-paradigm of dynamic routed models.

\newpage

\subsubsection*{Ethical Broader Impact Statement}
This paper introduces FastLane, a method to improve the scalability and efficiency of late-interaction retrieval models, specifically by enabling dynamic routing. Our contributions advance the state-of-the-art in retrieval by significantly reducing computational overhead while maintaining competitive performance metrics. The societal impact of this work is primarily centered on improving the accessibility and practicality of information retrieval systems, especially in large-scale applications such as search engines, recommendation systems, and question-answering platforms. By reducing latency and memory costs, this method can enable the deployment of more efficient systems, making advanced retrieval technology accessible to broader audiences, including resource-constrained settings. From an ethical standpoint, we note that retrieval models are not free from potential biases in their training data. While \namel~reduces computational costs, it inherits any biases present in the underlying data and models, which could impact its retrieval decisions. We encourage further research on bias mitigation in retrieval systems to enhance fairness and inclusivity. While there are potential societal consequences of this work, we do not identify specific risks or ethical concerns that require immediate mitigation beyond the aforementioned considerations.

\subsubsection*{Acknowledgments}
This work builds upon research initiated during a pre-doctoral internship at Google DeepMind. We are profoundly grateful to \href{https://rajeshjayaram.com}{Rajesh Jayaram}, \href{https://people.csail.mit.edu/mirrokni/Welcome.html}{Vahab Mirrokni}, \href{https://www.linkedin.com/in/majid-hadian-96a123a7/}{Majid Hadian}, \href{https://research.google/people/jasonlee/?&type=google}{Jason Lee}, and \href{https://www.cs.umd.edu/~laxman/}{Laxman Dhulipala} for their insightful suggestions and thoughtful clarifications on the MUVERA framework. We also thank \href{https://akmenon.github.io}{Aditya Menon}, \href{https://www.linkedin.com/in/hjaincs/}{Himanshu Jain}, \href{https://bsrinadh.github.io}{Srinadh Bhojanapalli}, \href{https://ankitsrawat.github.io/home/}{Ankit Singh Rawat}, \href{https://jiziwei.github.io}{Ziwei Ji}, \href{http://www.sanjivk.com}{Sanjiv Kumar}, and the broader BigML team at Google Research for their invaluable feedback, encouragement, and technical guidance throughout the project. Their collective expertise and support were instrumental in shaping this work. 

We are especially indebted to the vibrant and collaborative environment fostered by our colleagues at Google Research and Google DeepMind. In particular, we acknowledge \href{https://research.google/people/106704/?&type=google}{Manish Gupta}, \href{https://sites.google.com/view/karthikeyan-shanmugam}{Karthikeyan Shanmugam}, \href{https://www.cs.cmu.edu/~asuggala/}{Arun Sai Suggala}, \href{https://dheerajnagaraj.com}{Dheeraj Nagaraj}, \href{http://anshulmittal.org}{Anshul Mittal}, \href{https://www.cs.utexas.edu/~inderjit/}{Inderjit S. Dhillon}, \href{https://www.linkedin.com/in/nithi-gupta-7967b217/}{Nithi Gupta}, \href{https://www.linkedin.com/in/gausri108/}{Gaurav Srivastava} and the broader Google Research India team, whose engagement and insights greatly enriched this research project.

\bibliography{main}
\bibliographystyle{tmlr}

\newpage

\appendix

\section{Datasets}
\label{appendix:datasets}

In this section, we briefly discuss the open-source datasets used in this work adapted from a similar setup as \cite{menon2022defense}.

\paragraph{MS Marco} The MS Marco benchmark \citep{bajaj2016ms} has been included since it is widely recognized as the gold standard for evaluating and benchmarking large-scale information retrieval systems~\citep{thakur2021beir,ni2021large}. It is a collection of real-world search queries and corresponding documents carefully curated from the Microsoft Bing search engine. What sets MSMarco apart from other datasets is its scale and diversity, consisting of approximately 9 million documents in its corpus and 532,761 query-passage pairs for fine-tuning the majority of the retrievers. Due to the increased complexity in scale and missing labels, the benchmark is widely known to be challenging. The dataset has been extensively explored and used for fine-tuning dense retrievers in recent works \citep{thakur2021beir, nogueira2019passage, gao2020complementing, qu2020rocketqa}. MSMarco has gained significant attention and popularity in the research community due to its realistic and challenging nature. Its large-scale and diverse dataset reflects the complexities and nuances of real-world search scenarios, making it an excellent testbed for evaluating the performance of information retrieval algorithms.

\paragraph{TREC DL-19} The TREC Deep Learning 2019 (TREC DL-19) benchmark~\citep{craswell2020overview} is a well-established evaluation suite designed to assess the performance of retrieval systems in the context of deep learning-based search. It builds upon the MS MARCO passage ranking task by using a subset of its queries, enriched with high-quality relevance judgments from the TREC expert assessors. The benchmark comprises 43 queries derived from MS MARCO, with relevance annotations that are significantly more granular and reliable compared to automatic weak labels. TREC-DL 19 is particularly valued for its human-annotated labels and its emphasis on ranking quality, making it suitable for robust evaluation of fine-grained retrieval capabilities. Due to its curated annotations and widespread adoption in the IR community, the dataset is frequently used as a downstream evaluation benchmark to validate the real-world effectiveness of retrievers~\citep{thakur2021beir}. It serves as a critical diagnostic tool for testing generalization and transfer capabilities, especially in zero-shot and low-resource settings.

\section{Model}

\label{sec:models}

In line with prior work~\citep{menon2022defense}, we implement and train a diverse set of BERT-based retrieval architectures to evaluate the impact of static and dynamic routing strategies in multi-view document representations. 

\paragraph{Model Architecture.} 
For all retrievers, we use a 6-layer small-BERT encoder initialized from the pre-trained models of \citep{turc2019well}, with hidden size 768 and 12 attention heads. The document encoder produces contextualized token embeddings, while the query encoder either pools or scores these token-level embeddings depending on the retrieval architecture. All models are trained from scratch for 1.5 million steps until convergence, without using any large-scale in-domain pretraining or contrastive bootstrapping. For dual encoder models, query and document encoders are either shared (tied) or instantiated independently (untied), depending on the specific model variant. ColBERT-based models compute similarity via late interaction using a max or sum pooling over token-level inner products~\citep{santhanam2021colbertv2}.

\paragraph{Model Variants.} 
We predominantly compare the following retrieval architectures, but also compare against other State-of-the-Art models for the sake of completeness as depicted in Appendix~\ref{appendix:extended_results}. We list few of these model variants below:
\begin{itemize}
    \item \textbf{Dual-encoder BERT}: A baseline model with tied query and document encoders and a single [``CLS''] token used for computing dot-product similarity. This serves as the canonical dual encoder setup which is typically deployed in production.
    \item \textbf{ColBERT (max-max)}: A ColBERT-style model using max pooling on both query and document token representations.
    \item \textbf{Static-Tied}: A static routing baseline where the query and document encoders are tied. The first view (or ``CLS'' token) is always selected as the query representation, making it a single-view static retriever with minimal capacity.
    \item \textbf{Static-Split}: Similar to Static-Tied, but the query and document encoders are untied. This configuration increases model capacity slightly while retaining a static routing policy.
    \item \textbf{FastLane (ours)}: Our full model with learned dynamic routing over multiple query views. The model selects the most relevant view for each query through a learned router, trained end-to-end alongside the retriever. Query and document encoders are untied, and token-level matching is performed using a sum-aggregate scoring.
    \item \textbf{ColBERT (sum-max)}: A stronger late-interaction variant where token-level similarities are aggregated using sum over queries and max over documents, yielding the best overall performance.
\end{itemize}

\paragraph{Training Details.}
All models are trained with AdamW~\citep{loshchilov2017decoupled} using a linear warmup and cosine decay scheduler. We tune learning rates in the range $\{5 \times 10^{-6}, 1 \times 10^{-5}, 3 \times 10^{-5}, 5 \times 10^{-5}\}$ separately for each model variant to ensure fair comparisons. Weight decay is set to 0.01 across our experiments. Following~\citet{hofstatter2020improving}, we use KL-divergence based distillation from a strong BM25 teacher, which yields modest but consistent gains across all ColBERT-based variants as also reported by \citet{menon2022defense}. 

\section{Additional Results}
\label{appendix:extended_results}
Table~\ref{tab:main_comparison} depicts the main results of our experiments, comparing Fastlane to several SOTA baseline models.  As shown in the table, Fastlane achieves competitive performance across both the MS MARCO and TREC DL-19 datasets.  Specifically, Fastlane demonstrates notable improvements in MRR@10 and nDCG@10 compared to the Dual-encoder BERT and ColBERT (max-max) models. While Fastlane slightly underperforms ColBERT (sum-max) on both datasets, the difference is relatively small.  We believe this slight performance trade-off is justified by the substantial gains in retrieval speed, as detailed in Section~\ref{sec:results}.

\begin{table*}[htbp]
\centering
\small
\caption{Performance metrics of $MRR@10$ and $nDCG@10$ evaluated on the MS MARCO dev dataset and TREC DL-19 dataset. The best value for each metric is indicated in \textbf{bold}, and the second-best is marked with $\dag$.}
\label{tab:main_comparison}
\resizebox{\linewidth}{!}{%
\begin{tabular}{lccccc}
\toprule
& \multicolumn{2}{c}{\textbf{MSMARCO re-rank}} & \multicolumn{1}{c}{\textbf{TREC DL19 re-rank}} \\
\textbf{Model} & \textbf{MRR} & \textbf{nDCG}  & \textbf{nDCG} & {\#. parameters.}\\
\midrule
DPR \citep{thakur2021beir} & - & 0.177 & - & 110M (BERT-base)\\
BM25 (official) \citep{khattab2020colbert} & {0.167} & 0.228 & 0.501 & - \\
BM25 (Anserini) \citep{khattab2020colbert} & 0.187 & - & -& - \\
\texttt{cpt-text} S  \citep{neelakantan2022text} & 0.199 & - &- & 300M (cpt-text S) \\
\texttt{cpt-text} M \citep{neelakantan2022text} & 0.206 & - & - & 1.2B (cpt-text M)\\
\texttt{cpt-text} L  \citep{neelakantan2022text}& 0.215 & - & - & 6B (cpt-text L)\\
\texttt{cpt-text} XL \citep{neelakantan2022text} & 0.227 & - & - & 175B (cpt-text XL) \\
DSI  (Atomic Docid + Doc2Query + Base Model) \citep{tay2022transformer} & 0.260 & 0.3228 & - & 250M (T5-base)  \\
DSI  (Naive String Docid + Doc2Query + XL Model) \citep{tay2022transformer} & 0.210 & - & - & 3B (T5-XL)\\
DSI  (Naive String Docid + Doc2Query + XXL Model) \citep{tay2022transformer} & 0.165 & - & - &  11B (T5-XXL)\\
DSI  (Semantic String Docid + Doc2Query + XL Model) \citep{tay2022transformer} & 0.203 & 0.2786 & - & 3B (T5-XL)\\
CCSA \citep{lassance2021composite} & 0.289 & - & 0.583 &  110M (BERT-base)\\
HNSW \citep{malkov2018efficient} & 0.289 & - & - & - \\
RoBERTa-base + In-batch Negatives \citep{monath2023improving} & 0.242 & - & - &  123M (RoBERTa-base)\\
RoBERTa-base + Uniform Negatives \citep{monath2023improving} & 0.305 & - & - &  123M (RoBERTa-base)\\
RoBERTa-base + DyNNIBAL \citep{monath2023improving} & 0.334 & - & - &  123M (RoBERTa-base)\\
RoBERTa-base + Stochastic Negatives \citep{monath2023improving} & 0.331 & - & - &  123M (RoBERTa-base)\\
RoBERTa-base +Negative Cache \citep{monath2023improving} & 0.331 & - & - &  123M (RoBERTa-base)\\
RoBERTa-base + Exhaustive Negatives \citep{monath2023improving} & 0.345 & - & - &  123M (RoBERTa-base)\\

SGPT-CE-2.7B \citep{muennighoff2022sgpt} & - & 0.278 & - & 2.7B (GPT-Neo)\\
SGPT-CE-6.1B \citep{muennighoff2022sgpt} & - & 0.290 & - & 6.1B (GPT-J-6B)\\
SGPT-BE-5.8B \citep{muennighoff2022sgpt} & - & 0.399 & - & 5.8B (GPT-J)\\
doc2query \citep{khattab2020colbert} & 0.215 & - & - & -\\
DeepCT \citep{thakur2021beir} & 0.243 & 0.296 & - & 110M (BERT-base)\\
docTTTTquery \citep{khattab2020colbert} & 0.2077 & - & - & -\\
SPARTA \citep{thakur2021beir} & - & 0.351 & - & 110M (BERT-base)\\
docT5query \citep{thakur2021beir} & - & 0.338 & - & -\\
ANCE & 0.330 & 0.388 & - & -\\
RepCONC \citep{zhan2022learning} & 0.340 & - & - & 123M (RoBERTa-base) \\
Dual-encoder BERT (6-layer) \citep{menon2022defense}& 0.344 & 0.404 & 0.742 & 67.5M (small-bert) \\
Cross-attention BERT (12-layer) \citep{menon2022defense}&- & - & 0.749 &  110M (BERT-base)\\
DistilBERT + MSE \citep{menon2022defense} &- & - & 0.693 & 66M (distilBERT)\\
TAS-B \citep{thakur2021beir} & - & 0.408 & - & 110M (BERT-base) \\
GenQ \citep{thakur2021beir} & - & 0.408 & - &-\\
ColBERT (re-rank) \citep{khattab2020colbert} & 0.348 & - &  - & 110M(BERT-base) \\
ColBERT (end-to-end) \citep{khattab2020colbert} & 0.360 & 0.40 & - & 110M (BERT-base) \\
BM25 + CE \citep{thakur2021beir} & - & 0.413 & - & -\\
SPLADE (simple training) \citep{formal2022distillation} & 0.342 & - & 0.699 & 110M (BERT-base) \\
DistilBERT + Margin MSE \citep{menon2022defense} & - & - & 0.718 & 66M (distilBERT) \\
DistilBERT + RankDistil-B \citep{menon2022defense} & - & - & 0.708 & 66M (distilBERT) \\
DistilBERT + Softmax CE \citep{menon2022defense} & - & - & 0.726 & 66M (distilBERT) \\
DistilBERT + $M^3$SE \citep{menon2022defense} & - & - & 0.714 & 66M (distilBERT) \\
SPLADE + DistilMSE \citep{formal2022distillation} & 0.358 & - & 0.729 & 66M (distilBERT) \\
SPLADE + SelfDistil \citep{formal2022distillation} & 0.368 & - & 0.723 & 66M (distilBERT) \\
SPLADE + EnsembleDistil \citep{formal2022distillation} & 0.369 & - & 0.721 & 66M (distilBERT) \\
SPLADE + CoCondenser-SelfDistil \citep{formal2022distillation} & 0.375 & - & 0.73 & 66M (distilBERT) \\
SPLADE + CoCondenser-EnsembleDistil \citep{formal2022distillation} & {0.380} & - & 0.732 & 66M (distilBERT) \\
\midrule
\namel~\textbf{(Ours)}& 0.372 & {0.430}$^\dag$ & {0.754}$^\dag$ & 67.5M (small-bert) \\

ColBERT (sum-max) & 0.384$^\dag$ & \textbf{0.443} & \textbf{0.757} & 67.5M (small-bert) \\
\midrule

ColBERT-v2 \citep{santhanam2021colbertv2} & \textbf{0.397} & - & - & 110M (BERT-base) \\
\bottomrule
\end{tabular}
}
\end{table*}


\end{document}